\documentclass[preprint,12pt]{elsarticle}
\usepackage{graphicx}
\usepackage[figurename=Fig.,bf]{caption}
\usepackage{epsfig}
\usepackage{subfig}
\usepackage{amsmath}
\usepackage{physics}

\begin{document}
\begin{frontmatter}

\address[famu]{Florida A\&M University,Department of Physics,
Tallahassee FL,32307}

\title{Wave functions for a toroidal quantum dot in the presence of an axially symmetric magnetic field: transition from ring to bulk states as a function of aspect ratio}

\author [famu]{M. Encinosa\corref{cor1}}
\ead{mario.encinosa@famu.edu}
\author [famu]{J. {Williamson}}
\ead{johnny1.williamson@famu.edu}

\cortext[cor1]{Corresponding Author}

\begin{abstract}
A basis set expansion is employed to calculate spectra and eigenstates of charge carriers within a toroidal volume characterized by major radius $R$ and minor radius $a$ immersed in an azimuthally symmetric magnetic field. The angular variables appearing in the Schrodinger equation are eliminated by contour methods, yielding effective potentials that reduce computational time by a large factor. An approximation formula for the single particle spectrum is presented that allows efficient construction of the partition function necessary for rapid calculation of thermodynamic quantities. The heat capacity from as a function of torus aspect ratio and temperature is calculated. The transition from ring-type to bulk-type behavior of the heat capacity is presented.
\end{abstract}


\end{frontmatter}

\section{Introduction}

The potential (and in many cases realized) optical and transport applications of broad classes of fabricated mesoscale structures will require that their properties be modeled with fidelity. The possibility of studying structures with tailored geometries has become an active area of investigation\cite{geoelec1, geoelec2}.

Nanometer scale rings and tori may find use in nanoelectronics, quantum computing \cite{gold}, and as biosensors \cite{biosensor1, biosensor2, biosensor3}. In particular, metallic tori and toroidal carbon nanotubes have been envisioned as components of nano-traps for cold polar molecules \cite{gold}, as circuit elements \cite{nanotorus1, nanotorus2}, as nano-photonic resonators \cite{optical1, plasmonic1} and as generators of novel electromagnetic moments \cite{tormoment1,encinosadipole, toroidalhelix}. In principle, magnetic flux through a conducting ring or torus can induce persistent currents, and attachment to metallic leads could provide a means of generating poloidal modes which give rise to toroidal moments \cite{tormoment1}. These effects would be enhanced by increasing the number of charge carriers available to participate in these processes.

This work concerns charged particles constrained to a toroidal region. For many applications \cite{gold,optical1} the use of classical physics is appropriate and sufficient when attempting to model such systems. This work  anticipates  circumstances where  a quantum mechanical description is required, i.e., situations in which analytic expressions for wave functions  may be of importance. To date, most efforts to model genus one structures have focused on flat rings, for which solutions to the Schr{\"o}dinger equation can be written in closed form -- this is not the case for the torus.

In \cite{bhks}, the Schrodinger equation for a single spinless charge carrier in a toroidal volume was developed by standard methods and solved using finite element methods on a disk within the structure. Azimuthal symmetry then allowed for the solution to be known everywhere. Here, a basis set expansion is introduced that renders several of the terms of the resulting Hamiltonian matrix $H_T$, diagonal in the basis.  The remaining terms are easily handled by standard and contour integrations. The contribution of each term in $H_T$ to a particular result is then easily determined as a function of applied flux. Furthermore, the nature of the solutions make it clear that a useful approximation can be developed for calculations in which large numbers of eigenvalues are required.

This paper is organized as follows: To keep this work self-contained, section 2 briefly develops the Hamiltonian $H_T$ for a particle in a symmetric torus. The basis set is presented, and a discussion of how a term within $H_T$ not separable into radial and angular variables can be reduced to an effective radial potential. Section 3 presents calculated spectra, eigenfunctions, and studies heat capacity as a function of aspect ratio $\alpha = a/R$.  Section 4 is reserved for discussion and conclusions.

\section{Basis set and Hamiltonian}
Points within a toroidal geometry characterized by a  major radius $R$ and minor radius $a$ are described by the Monge form
\begin{equation}
{\bf r}(\rho,\theta,\phi)=W(\rho,\theta)\,\hat{\rho}+\rho\,{\rm sin}\theta\,\hat{\textrm{k}},
\end{equation}
\noindent where
\begin{equation}
W(\rho,\theta)=R+\rho\,{\rm cos}\theta.
\end{equation}
\noindent In the Coulomb gauge, the Schr\"{o}dinger equation becomes
\begin{equation}
\bigg( \nabla ^2 + 2{\textit i} \frac{e}{\hbar}~{\bf A} \cdot \vec{\nabla} - \frac{e^2}{\hbar^2} {\bf A}^2 +  \frac{2 {\textit m}_e E}{\hbar^2}  \bigg) \Psi = 0,
\end{equation}
where $e$ is the electron charge.
\noindent The gradient found by standard methods (the unit vectors are listed in appendix (a)) is
\begin{equation}
\hat{{n}} \frac {\partial} {\partial \rho}+\frac{1}{\rho} \hat{\theta}\frac{\partial}{\partial \theta}+ \frac{1}{W(\rho,\theta)} \hat{\phi}\frac{\partial}{ \partial \phi}.
\end{equation}


\vskip 8pt
\noindent
Upon defining
$$u = \frac{\rho}{R}$$
$$\gamma_0 = B \pi R^2$$
$$\gamma_{\mbox{\tiny N}} = \frac{\pi \hbar}{e}$$
$$\epsilon = \frac{2 {\textit m}_e E R^2}{\hbar^2}$$
$$\tau_0 = \frac{\gamma_0}{\gamma_{\mbox{\tiny N}}}$$
$$W(u,\theta)=1+u \cos\theta,$$
\noindent Eq. [3] takes dimensionless form
\begin{equation}
\begin{split}
\bigg[\frac{\partial^2}{\partial u^2} + & \frac{1}{u}\frac{\partial}{\partial u}  + \frac {1} {u^2} \frac{\partial^2}{\partial \theta^2}+  \frac {\cos\theta} {W(u,\theta)} \frac{\partial}{\partial u} + \frac {1}{W^2 (u,\theta)} \frac{\partial^2}{\partial \phi^2}  \\ & -\frac {\sin\theta} {W(u,\theta)} \frac {\partial} {\partial \theta}+  \frac {\tau_0} {i} \frac {\partial}{\partial \phi} - \frac {1} {4} \tau^2_0 W^2(u,\theta) + \epsilon \bigg]\psi = 0.
\end{split}
\end{equation}
\vskip 8pt
\noindent
It is customary to define an auxiliary function
\begin{equation}
\psi(\rho,\theta,\phi)=\frac {1} {\sqrt{W(\rho,\theta)}} \chi(\rho,\theta)\exp[i m_\phi \phi]
\end{equation}
\vskip 8pt
\noindent
which allows Eq.[5] to be expressed after insertion of Eq.[6] as
\begin{equation}
\bigg[\frac{\partial^2}{\partial u^2} + \frac{1}{u}\frac{\partial}{\partial u} +  \frac {1} {u^2} \frac{\partial^2}{\partial \theta^2} + \frac {(\frac{1} {4}-m_\phi^2)}{W^2 (u,\theta)} +  {\tau_0} {m_\phi} - \frac {1} {4} \tau^2_0 W^2(u,\theta) + \epsilon \bigg]\chi(\rho,\theta)=0.
\end{equation}
\vskip 8pt
\section{Computational methods}
The first three terms in Eq.[7] comprise part of Bessel's equation, and the $\theta \rightarrow -\theta$ invariance of the Hamiltonian allows its solutions to be separated into even and odd basis functions. Therefore, a natural choice of basis with a hard wall boundary condition at $u = \alpha$, is
\begin{equation}
\chi{^{\pm}_{m_\phi}} (\rho,\theta)=\sum_{n\nu}\frac {1}{N_{n\nu} T_n} C{^{n\nu}_{m_\phi}}J_n \biggl (\frac{x_{n\nu}u} {\alpha} \biggr)\begin{pmatrix}\cos n\theta \\ \sin n\theta \end{pmatrix}
\end{equation}
\noindent with $N_{n\nu}$ and $T_n$ the standard Bessel and trigonometric normalization factors.

The only term in $H_T$ not trivially separable into a product of functions in $u$ and $\theta$ can be integrated by contour methods to yield effective potentials $\Lambda_{\bar{n} n}^{\pm}(u)$. The relevant matrix elements are
\begin{equation}
\begin{split}
H_{\bar{n}\bar{\nu}, n \nu} =   \bigg [ -\biggl (\frac {x_{n\nu}} {\alpha}\biggr )^2 +m_\phi \tau_0 - \frac {\tau_0^2} {4} \bigg ] \delta_{\bar{n}\bar{\nu} n \nu} -  \\  \bra{\bar{n}\bar{\nu}} (\frac{1} {4}-m_\phi^2) \Lambda_{\bar{n} n}^{\pm}(u)+  \frac {1} {2}\tau^2_0 u  f^{\pm}_1(n ,\bar{n}) + \frac {1} {8}\tau^2_0 u^2  f^{\pm}_2(n, \bar{n})\ket{{n \nu}}
\end{split}
\end{equation}
\noindent where
\begin{equation}
f^{\pm}_k(\bar{n} ,n) \equiv \int_{0}^{2\pi} d\theta  \cos \bar{n}\theta \begin{pmatrix}\cos k \theta \\ \sin k \theta \end{pmatrix} \cos n\theta.
\end{equation}
\noindent Explicit forms of $\Lambda_{\bar{n} n}^{\pm}(u)$ are presented in the appendix (b).

\section{Results}
Representative eigenvalues and eigensolutions resulting from diagonalization of Eq.[7] are presented in tables [1-2] for the positive parity states; the negative parity states exhibit similar behavior.  With $\alpha = 0.4$, the  diagonalized $| n \nu m_\phi>_D $ state is well approximated by its corresponding basis state for lower $m_\phi$ values, with admixtures of higher states becoming non-negligible for $m_\phi \sim 8$.  As $\alpha$ is decreased the admixture coefficients are substantially reduced. In neither case does a $1T$ field significantly alter the result.

\begin{table}
\centering
\resizebox{\columnwidth}{!}{
\begin{tabular}[t]{l l}
$\alpha=0.4$&$|11,0\rangle_D$ state superposition.\\ \hline
$m=0$& $|11,0\rangle_D \approx |11,0\rangle +O(10^{-4})  $ \\
$m=5$& $|11,5\rangle_D \approx (-.987|11,5\rangle +.119|01,5\rangle+.090|21,5\rangle) e^{i5\phi} $\\
$m=10$& $|11,10\rangle_D \approx (.854|11,10\rangle -.361|01,10\rangle+ .318|21,10\rangle) e^{i10\phi} $\\
\hline
\end{tabular}
}
\caption{Representative behavior of a basis state as $m_\phi$ is increased.}
\label{tab1}
\end{table}

\begin{table}
\centering
\resizebox{\columnwidth}{!}{
\begin{tabular}[t]{l l}
$\alpha=0.2$&$|11,0\rangle_D$ state superposition.\\ \hline
$m=0$& $|11,0\rangle_D \approx |11,0\rangle +O(10^{-5})  $\\
$m=5$& $|11,5\rangle_D \approx ( |11,5\rangle +O(10^{-3})) e^{i5\phi} $\\
$m=10$& $|11,10\rangle_D \approx (.997|11,10\rangle -.059|01,10\rangle+ .041|21,10\rangle) e^{i10\phi} $\\
\hline
\end{tabular}
}
\caption{Representative behavior of a basis state as $m_\phi$ is increased. Reduction of the value of $\alpha$ reduces the admixtures substantially.}
\label{tab2}
\end{table}

Tables [3-5] illustrate how the eignenvalues obtained from diagonalizing Eq.[7] in comparison to approximate values obtained using
\begin{equation}
\varepsilon(n, \nu, m_\phi, \gamma) =\biggl ( \frac {x_{n \nu}} {\alpha}\biggr )^2 + m^2_{\phi}- \frac {1} {4} + \gamma m_\phi + \frac {\gamma^2} {4},
\end{equation}
where $x_{n\nu}$ are the zeros of Bessel functions of the first kind
for $\alpha = 0.4$ (a survey of images of fabricated structures generally show tori having aspect rations well below this value). It is simple to further refine the formula to increase accuracy using perturbative methods should there be a requirement for higher accuracy at larger $(n, \nu)$.

\begin{table}
\centering
\resizebox{\columnwidth}{!}{
\begin{tabular}[t]{c c c c c c c}
\hline
$m_{\phi}=0$ &\multicolumn{2}{c}{$B=0.0 \,\textrm{T}$}&\multicolumn{2}{c}{$B=0.5\,\textrm{T}$}&\multicolumn{2}{c}{$B=1.0\,\textrm{T}$}\\
\cline{2-7}
$\alpha = 0.4$ &Exact&\multicolumn{1}{c}{Approx.} &Exact&\multicolumn{1}{c}{Approx.}&Exact&Approx.\\ \hline
$E_{01}$& 35.88&35.89 & 36.01 & 36.03 & 36.41& 36.42\\
$E_{11}$& 91.48&91.51 & 91.62 & 91.64 & 92.03& 92.04\\
$E_{21}$& 164.56& 164.59 &164.70 & 164.72&165.11& 165.12\\
$E_{02}$& 190.57& 190.20 &190.30 & 190.33 &190.71& 190.72\\
$E_{12}$& 307.33& 307.37 &307.47 & 307.50& 307.88& 307.89\\
$E_{22}$& 442.54& 442.56 &442.67 & 442.69& 443.08& 443.09\\
$E_{03}$& 467.77& 467.79 &467.91 & 467.93& 468.31& 468.32\\
$E_{13}$& 646.59& 646.62 &646.72 & 646.75& 647.13& 647.15\\
$E_{23}$& 843.61& 843.63 &843.74 & 843.76& 844.17& 844.16\\
\hline
\end{tabular}
}
\caption{$m_\phi = 0$ basis set and approximate eigenvalues at $B_0$ = .0, .5, 1 T, $\alpha =0.4$.}
\label{tab3}
\end{table}

\begin{table}
\centering
\resizebox{\columnwidth}{!}{
\begin{tabular}[t]{c c c c c c c}
\hline
$m_{\phi}=5$&\multicolumn{2}{c}{$B=0.0 \,\textrm{T}$}&\multicolumn{2}{c}{$B=0.5\,\textrm{T}$}&\multicolumn{2}{c}{$B=1.0\,\textrm{T}$}\\
\cline{2-7} $\alpha = 0.4$
&Exact&\multicolumn{1}{c}{Approx.}&Exact&\multicolumn{1}{c}{Approx.}&Exact&Approx.\\ \hline
$E_{01}$& 61.42& 60.89 & 65.18 & 64.64    & 69.24 & 68.67\\
$E_{11}$& 119.74& 116.51 & 123.49 & 120.26 & 127.54& 124.29\\
$E_{21}$& 192.71& 189.59 &196.45 & 193.34 &200.48& 197.38\\
$E_{02}$& 217.28& 215.20 &221.03 & 218.93 &225.071& 222.97\\
$E_{12}$& 335.79& 332.37 &339.53 & 336.12 & 343.57& 340.14\\
$E_{22}$& 470.10& 467.56 &473.84 & 471.31 & 477.88& 475.34\\
$E_{03}$& 495.06& 492.79 &498.81 & 496.55 & 502.85& 500.57\\
$E_{13}$& 675.32& 671.62 &678.98& 675.37  & 683.02& 679.40\\
$E_{23}$& 871.19& 868.63 &874.94 & 872.38 & 878.98& 876.41\\
\hline
\end{tabular}
}
\caption{$m_\phi = 5$ basis set and approximate eigenvalues at $B_0$ = .0, .5, 1 T, $\alpha =0.4$.}
\label{tab4}
\end{table}

\begin{table}
\centering
\resizebox{\columnwidth}{!}{
\begin{tabular}[t]{c c c c c c c}
\hline
$m_{\phi}=10$&\multicolumn{2}{c}{$B=0.0 \,\textrm{T}$}&\multicolumn{2}{c}{$B=0.5\,\textrm{T}$}&\multicolumn{2}{c}{$B=1.0\,\textrm{T}$}\\
\cline{2-7} $\alpha = 0.4$
&Exact&\multicolumn{1}{c}{Approx.}&Exact&\multicolumn{1}{c}{Approx.}&Exact&Approx.\\ \hline
$E_{01}$& 130.22 & 135.90 & 137.61 & 143.27 & 145.36 & 150.92\\
$E_{11}$& 202.07& 191.51 & 209.43 & 198.88 & 217.12& 206.54\\
$E_{21}$& 278.18& 264.51 & 285.53 &271.96 & 293.19 & 279.62\\
$E_{02}$& 301.12& 290.20 &308.49 & 297.57 & 316.15& 305.22\\
$E_{12}$& 421.44& 407.37 & 336.12 & 343.57& 436.49&  422.39\\
$E_{22}$& 552.67& 542.56 &428.81 & 414.74 & 567.69& 557.59\\
$E_{03}$& 578.70& 567.79 &560.03  &575.15 & 593.74  & 582.82\\
$E_{13}$& 763.58& 746.62 &770.94 & 753.99  & 778.60 & 761.65\\
$E_{23}$& 956.26& 943.63 &963.62 &951.00 &971.28 & 958.66\\
\hline
\end{tabular}
}
\caption{$m_\phi = 10$ basis set and approximate eigenvalues at $B_0$ = .0, .5, 1 T, $\alpha =0.4$.}
\label{tab5}
\end{table}

Approximating the eigenvalues produced from Eq.[9] by $\varepsilon(n,\nu, m_\phi, \gamma)$ simplifies calculation of expectation values, and the approximate eigenvalues allow efficient calculation of thermodynamic quantities. For example, the partition function
\begin{equation}
Z(\tau)=\sum_N \exp[-{E_N}/{\tau}]
\end{equation}
\noindent was calculated with fewer than $50$ states in  \cite{bhks} at 77 K and 300 K for a presumably $R=308$ \AA, $\alpha = 0.4$ torus where GaAs parameters for the effective mass and dielectric constant set the distance and energy scales. The partition function in \cite{bhks} at 77K is reproduced, Z(300K) is not. For 50 states,\cite{bhks} showed convergence to 1.91, whereas this work determined Z(300)=3.19 for 50 states via diagonalization, and to 3.21 with a large number of approximated eigenvalues. What parameters were used to generate the corresponding results in Table 1 in \cite{bhks} were not explicitly stated and may differ from what has been assumed here.

With $Z(\tau)$ available, thermodynamic quantities can be determined, and several were presented in \cite{bhks}. This work focuses on computing heat capacity as a thermodynamic means of exploring the transition from ring-like to bulk-like behavior of nanoscale tori. Standard analysis yields $C_v = 1/2$ (dimensionless) for the ring and $C_v = 3/2$ (dimensionless) for an ideal gas. In Fig.[1] the heat capacity as a function of $\alpha$ is plotted for several temperatures. At lower temperatures, the from one to three degrees of freedom does not occur independent of the value of $\alpha$. As the temperature is increased, the transition occurs at a lower value of  $\alpha$ as expected. The behavior of specific heat as a function of temperature for several $\alpha$ illustrates the ring-to-bulk transition from a different perspective is shown in Fig.[2]. It is more apparent in this plot where the onset of the transition to two degrees of freedom occurs for a given $\alpha$; given the Bessel function nodal structure, it is likely that the $\theta$ modes are being activated first.
\begin{figure}
\centering
  \includegraphics[width=1.0\columnwidth]{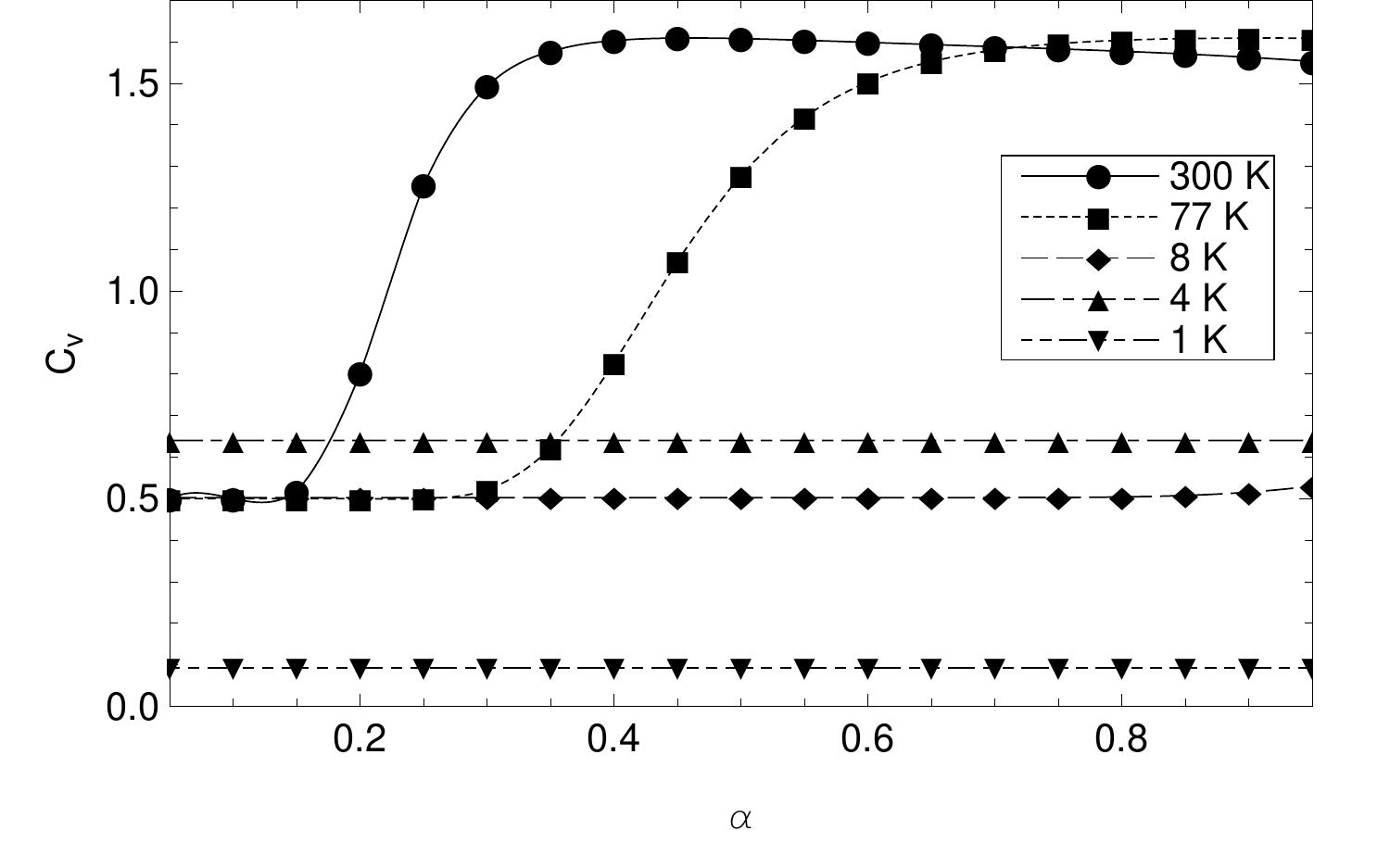}
  \caption{Heat capacity as a function of aspect ratio $\alpha$ for several temperatures.}
  \label{fig:Fig1final}
\end{figure}
\begin{figure}
\centering
  \includegraphics[width=1.0\columnwidth]{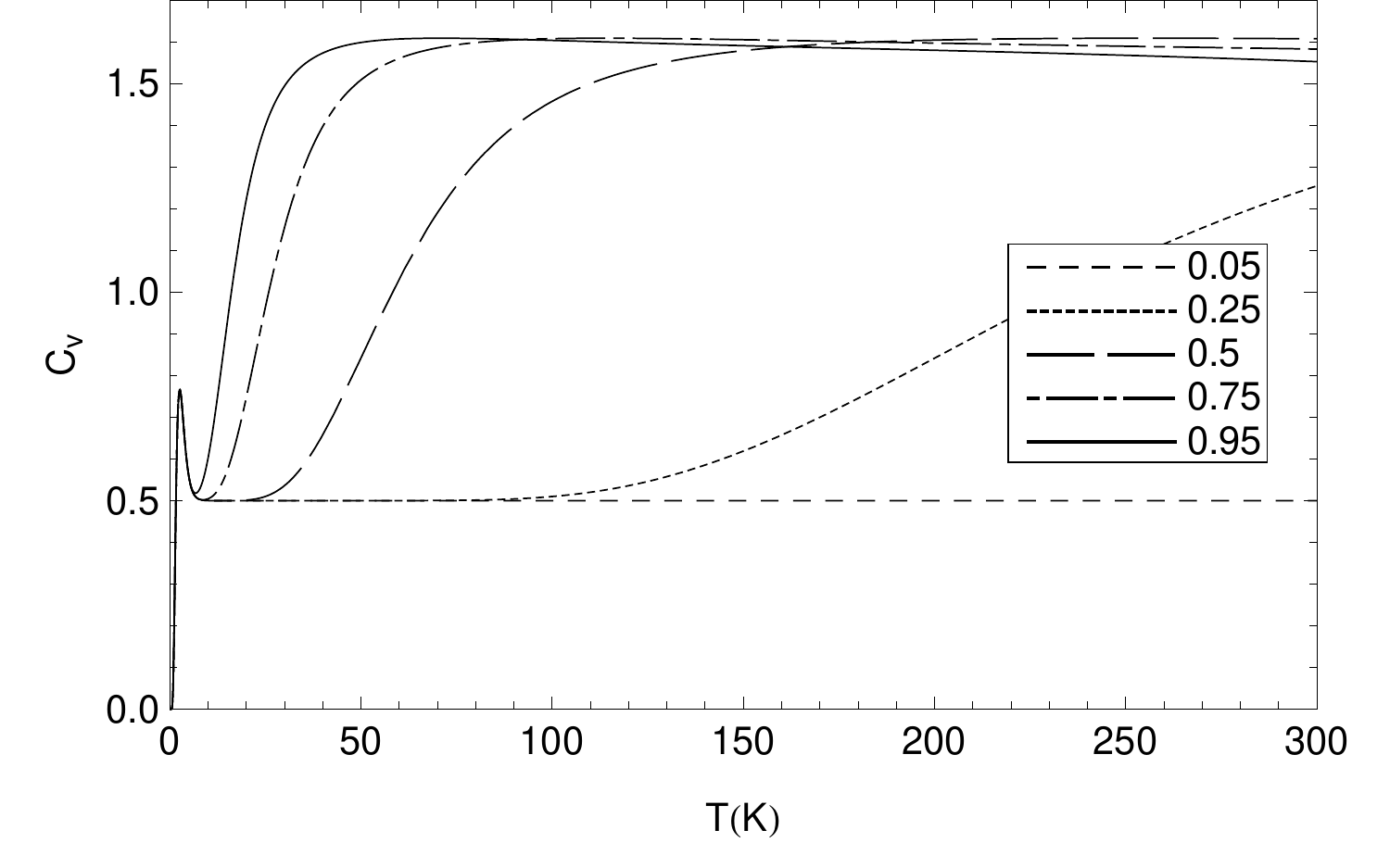}
  \caption{Heat capacity as a function of temperature in Kelvin for several $\alpha$.}
  \label{fig:Fig2final}
\end{figure}
\section{Conclusions}
In this work a basis set expansion in Bessel and trigonometric functions was employed to calculate the spectrum and eigenfunctions of a charged particle in a toroidal volume as functions of geometry and applied magnetic field. It was shown that a single basis term is sufficient for many low-lying states is sufficient to approximate a solution comprising several basis terms. Should there be a need for increased accuracy, other states in the expansion are easily included.  It should also be noted that integrating out the $\theta$ degree of freedom reduces the processor time required to diagonalize Eq.[7] in \textit{Mathematica} on a desktop PC by a factor of approximately 400 for the 9-state system considered here.

One obvious advantage of having analytic forms arises when considering the possibility of the structure being subject to an azimuthally dependent potential (for example, if the structure is attached to electrical leads). The basis set formulation presented here allows simple extension to such cases, particulary if the potential is periodic. The methods by which they can be included was developed in \cite{toroidalhelix}.

With solutions in hand, the calculations performed in \cite{bhks} and here involved assigning GaAs carrier parameters to the particles, a treatment parallel to what would be performed for a free electron gas.  However, the extension of this work to calculating electromagnetic properties is complicated by intrinsic GaAs possessing very few carriers in the conduction band, and those carriers must be described by the Fermi-Dirac distribution that satisfy the Heaviside $\mathcal{H}(x)$ condition
\begin{equation}
N =\sum \frac {\mathcal{H}( \beta \varepsilon(n,\nu,m_\phi,\gamma)-E_g/2)}  {1 + \exp[\beta \varepsilon(n,\nu,m_\phi,\gamma)]}
\end{equation}
\noindent with $\beta = \hbar^2/(2m^*_e R^2)$, and $E_g$ the band gap.  The accepted value of $E_g = 1.42 \,eV$; in practice, $E_g/2  \sim .7 \,eV$ inserted in Eq. [EE] gives a carrier concentration in good agreement with the accepted value of $2.0(10^6)/cm^3$. It is immediately evident that averaging electromagnetic operators over Eq.[13] gives rise to experimentally inaccessible values. A more promising approach involves investigation of either doped GaAs or metallic (specifically gold) nanotori, such as one described in \cite{gold} with $R= 875$, $\alpha =0 .29$ and $E_F \sim 5.5 \,eV$.  Preliminary work has established the Fermi surface intercepts are $(n, \nu, m_\phi) = (400, 80, \pm 1500)$; values for which an approximation formula proves indispensable.

Besides those noted  above, there are other natural extensions of this work. The expansion provided in Eq.[8] can be carried over to a shell of inner radius $b < a$ by allowing
\begin{equation}
J_n \bigg( \frac {x_{n\nu}u} {\alpha}\bigg)\rightarrow J_n (\eta_{n\nu} \alpha )Y_n (\eta_{n\nu}u)-J_n(\eta_{n\nu}u) Y_n ( \eta_{n\nu} \alpha)
\end{equation}
\noindent
with $Y_{n}$ a Bessel function of the second kind. Previous work regarding solutions on a toroidal surface indicate the angular dependence of those eigenfunctions will be more complicated \cite{scriptaenc,encinosaT2}.

Another possibility is to consider a Green's function formulation via
\begin{equation}
G({\bf r},{\bf{r}}^{\prime})=\sum_{all} \frac {1}  {2\pi N^2_{n\nu}
T^2_n }
J_n \big( \frac {x_{n\nu}u} {\alpha}\big ) J_n \big( \frac {x_{n\nu}u^\prime} {\alpha}\big )
e^{im_\phi(\phi-\phi^\prime)} \cos n\theta \cos n \theta^\prime + \, {\text{sin terms}}
\end{equation}
\noindent that allows calculation of the density of states, transmissivity, and conductivity of the torus. It is conceivable that coupling the torus to leads will provide sufficiently large imaginary contributions to the $\theta$-components of the Green's function, such that poloidal modes (if produced) could give rise to toroidal moments.
\section{Appendix}
\noindent a. Explicit form for the unit vectors in this work are
\begin{equation}
\hat{n}= \cos\theta \,\hat{\rho} + \sin \theta \,\hat{\textrm{k}}
\end{equation}
\begin{equation}
\hat{\theta} = -\sin\theta \,\hat{\rho} + \cos \theta \,\hat{\textrm{k}}
\end{equation}
\begin{equation}
\hat{\phi}= -\sin \phi \,\hat{\textrm{i}} + \cos \phi \,\hat{\textrm{j}}\,.
\end{equation}
\vskip 8pt
\noindent b. Only one term in $H_T$ is not trivially separable in the $u$ and $\theta$. The even and odd parity expansions require respectively
\begin{equation}
\Lambda_{\bar{n} n}^{+}(u) = \int_{0}^{2\pi} d\theta \frac { \cos \bar{n}\theta \cos \theta  \cos n\theta}
{(1 + u\cos\theta)^2}
\end{equation}
\begin{equation}
\Lambda_{\bar{n} n}^{-}(u) = \int_{0}^{2\pi} d\theta \frac { \cos \bar{n}\theta \sin \theta  \cos n\theta}
{(1 + u\cos\theta)^2}.
\end{equation}
Let $w(u)=\sqrt{1-u^2}$ and define
\begin{equation}
I_1 =\frac {1} {[w(u)]^{3/2}} \big [- \frac {1} {u}(1 - w(u)) \big ]^{\bar{n}+n} \big [( {\bar{n}+n})w(u)+1 \big ]
\end{equation}
\begin{equation}
I_2 =   \frac {1} {[w(u)]^{3/2}} \big [-\frac {1} {u}( 1 - w(u)) \big ]^{|{\bar{n}-n}|} \big [( |{\bar{n}-n})|w(u)+1 \big ].
\end{equation}
\noindent Then
\begin{equation}
\Lambda_{\bar{n} n}^{\pm}(u) = \pi (I_2 \pm I_1).
\end{equation}

\section{Acknowledgements}
M.E. would like to thank Lewis Johnson for providing software support for this work.
\bibliographystyle{ieeetr}
\bibliography{referencebase}

\begin{thebibliography}{10}

\bibitem{geoelec1}
A.~Tavkhelidze {\em IOP Conf. Ser.: Mater. Sci. Eng.}, vol.~60, p.~012055,
  2014.

\bibitem{geoelec2}
X.~Tang {\em J Phys Condensed Matter}, vol.~23, pp.~0953--8984, 2014.

\bibitem{gold}
M.~Salhi, A.~Passian, and G.~Siopsis {\em Phys. Rev. A}, vol.~92, p.~033416,
  2015.

\bibitem{biosensor1}
J.~R. Li, L.~Shi, Z.~Deng, S.~H. Lo, and G.~yu~Liu {\em Biochemistry}, vol.~51,
  p.~5876–5893, 2012.

\bibitem{biosensor2}
J.~R. Li, L.~Shi, Z.~Deng, S.~H. Lo, and G.~yu~Liu {\em Acc. Chem. Res.},
  vol.~46, pp.~2888--2897, 2013.

\bibitem{biosensor3}
Interuniversity Microelectronics Centre (IMEC). ScienceDaily.
  www.sciencedaily.com/releases/2010/06/100623085841.htm, 2010.

\bibitem{nanotorus1}
M.~Kowsar and H.~Sabzyan {\em Molecular Physics}, vol.~DOI:
  10.1080/00268976.2018, p.~150374, 2018.

\bibitem{nanotorus2}
L.~Liu, F.~Liu, and J.~Zhao {\em Nano Research}, vol.~7, p.~150374, 2014.

\bibitem{optical1}
F.~Beuerle, C.~Herrmann, A.~C. Whalley, C.~Valente, A.~Gamburd, M.~A. Ratner,
  and J.~F. Stodda {\em Chem. Eur. J.}, vol.~17, p.~3868–3875, 2011.

\bibitem{plasmonic1}
D.~W. Watson, S.~D. Jenkins, J.~Ruostekoski, V.~A. Fedotov, and N.~I. Zheludev
  {\em Phys. Rev. B}, vol.~93, p.~125420, 2016.

\bibitem{tormoment1}
N.~A. Spaldin, M.~Fiebig, and M.~Mostovoy {\em Journal of Physics: Condensed
  Matter}, vol.~20, p.~434203, 2008.

\bibitem{encinosadipole}
M.~Encinosa and M.~Jack {\em Journal of Computer-Aided Materials Design},
  vol.~14, pp.~65--71, 2007.

\bibitem{toroidalhelix}
J.~Williamson and M.~Encinosa. arXiv:1108.5097.

\bibitem{bhks}
D.~Baghdasaryana, D.~Hayrapetyana, E.~Kazaryana, and H.~Sarkisyanabc {\em
  Physica E}, vol.~101, pp.~1--4, 2018.

\bibitem{scriptaenc}
M.~Encinosa, L.~Mott, and B.~Etemadi {\em Physica Scripta}, vol.~72,
  pp.~13--15, 2005.

\bibitem{encinosaT2}
M.~Encinosa {\em Physica E}, vol.~28, pp.~209--218, 2005.

\end{thebibliography}

\end{document}